\documentclass[prl,twocolumn,showpacs,superscriptaddress,amsmath,amssymb,floatfix]{revtex4}
\usepackage{graphicx}
\usepackage{dcolumn}
\usepackage{bm}
\usepackage{amsmath}
\usepackage{latexsym}
\usepackage{amsfonts}
\usepackage{amssymb}
\usepackage[rightcaption]{sidecap}
\usepackage{float}
\usepackage{color}
\usepackage{soul}

\makeatletter

\newcommand{\Rmnum}[1]{\expandafter\@slowromancap\romannumeral  #1@}
\makeatother

\begin{document}
\title{Time Dependent Density Functional Theory meets Dynamical Mean Field Theory:  Real-Time  Dynamics
for the 3D Hubbard model}
\date{\today}
\author{Daniel Karlsson}
\affiliation{Mathematical Physics and European Theoretical Spectroscopy Facility (ETSF), Lund University, 22100  Lund, Sweden}
\author{Antonio  Privitera}
\affiliation{Institut f\"ur Theoretische Physik, Johann Wolfgang Goethe-Universit\"at, 60438 Frankfurt am Main, Germany}
\author{Claudio Verdozzi}
\affiliation{Mathematical Physics and European Theoretical Spectroscopy Facility (ETSF), Lund University, 22100  Lund, Sweden}
\begin{abstract}
\noindent We introduce a new class of exchange-correlation potentials for a static and time-dependent Density Functional Theory of 
strongly correlated systems in 3D. The potentials are obtained via Dynamical Mean Field Theory and, for strong enough interactions,
exhibit a discontinuity at half filling density, a signature of the Mott transition. 
For time-dependent perturbations, the dynamics is described in the adiabatic local density approximation.
Results from the new scheme compare very favorably to exact ones in clusters. As an application, we study Bloch
oscillations in the 3D Hubbard model.
\end{abstract}
\pacs{71.10.-w, 71.27.+a, 31.70.Hq, 71.10. Fd}

\maketitle
%

Time-dependent quantum phenomena hold an important place in today's condensed matter research.
A major theoretical challenge in this field is to describe strongly correlated systems out of  equilibrium.

In the last decade, Time-Dependent Density Functional Theory (TDDFT) has 
gained favor as a computationally viable, in principle exact time-dependent description of materials
\cite{rg84, TDDFTbook}. The basic TDDFT variable is the one-particle density $n$ and 
a key ingredient is the time-dependent exchange-correlation potential $v_{xc}$, embodying 
the complexities of the many-body problem. TDDFT applied to strongly correlated systems is in its beginnings.
Describing these systems in equilibrium with static density functional theory (DFT) \cite{staticDFT}
is already a difficult task \cite{DFTproblems}. TDDFT retains these difficulties, but also adds another hurdle:
Since time enters explicitly the formulation, $v_{xc}$ depends on the history of $n$ (memory effects) \cite{rg84,TDDFTbook}. 

In equilibrium, an effective {\it ab-initio} method to describe strong correlations is the LDA+DMFT
\cite{LDA+DMFT1,LDA+DMFT2}, combining DFT in the local density approximation (LDA) with Dynamical Mean Field Theory (DMFT)\cite{metzner_vollhardt}.
DMFT, which treats correlations nonperturbatively  via a local self-energy $\Sigma$ \cite{revdmft}, 
is also at the core of the DMFT+GW \cite{DMFTGW, Karlsson}, another
{\it ab-initio} method, which deals with nonlocal correlations within
the GW approximation \cite{Hedin}. 
These DMFT-based methods rely on Green's function formulations, and
the practical feasibility (in a foreseeable  future) of a nonequilibrium generalization
is not easy to assess, since Green's-function propagation
scales quadratically \cite{RvanLeeuwen, Freericks, Puig} with the simulation time.

TDDFT dynamics involves only one time variable.
It would thus be useful to have exchange-correlation potentials suitable
for strongly correlated systems. In equilibrium, they could offer a better start 
for Green's function based {\it ab-initio} schemes. Out of equilibrium, 
they could be used for adiabatic LDA \cite{ALDA} dynamics via TDDFT
and possibly be improved by including memory effects, absent in the adiabatic LDA. 

In this Letter we suggest a novel avenue to deal with strongly correlated systems in 3D and out of equilibrium, 
by combining DMFT with TDDFT.  For model strongly correlated systems in 1D, 
exchange-correlation potentials for DFT were introduced \cite{Capelle03,BAspin}, and a Bethe-Ansatz-based LDA (BALDA)
for $v_{xc}$ was proposed.
Such $v^{BALDA}_{xc}$ was then used to introduce an adiabatic scheme for the TDDFT of the 1D Hubbard model  \cite{Verdozzi08}, 
and the spin dependent case was considered in \cite{Polini08}.  
However, some interesting effects due to correlations are specific to 3D materials \cite{DFTproblems}.
This requires new exchange-correlation potentials for [TD]DFT, that we propose here to obtain via Dynamical Mean Field Theory.

We illustrate our method using the inhomogeneous 3D Hubbard model; as an initial application, we look at the Bloch oscillations
in this model. Our main findings are: i) for the homogeneous 3D Hubbard model, above a critical interaction $U_c^{Mott}$, $v_{xc}$
becomes a discontinuous function of $n$ at half-filling (the size of the discontinuity increases at larger $U$:s):
this is how the Mott metal-insulator transition manifests itself in $v_{xc}$; 
ii) the time-dependent densities from TDDFT-DMFT in the adiabatic LDA (hereafter referred to as $A^{LDA}_{DMFT}$) compare
very well with the exact ones in clusters; the agreement deteriorates for significantly non-adiabatic/strong 
perturbations; iii) $A^{LDA}_{DMFT}$ gives a good description of the correlation induced beats in the Bloch oscillations, but
no clear signatures of the damped regime, a fact most likely due to the lack of non-adiabaticity
in our exchange-correlation potentials. While explicit for the 3D Hubbard model, our results also provide insight
into the scope of TDDFT-DMFT for real strongly correlated materials.


{\it The Model.-} The time-dependent Hamiltonian for the Hubbard model is
\begin{align}
\!\!\!\!\hat{H}(\tau)\!=\!-t\!\!\sum_{\langle ij\rangle,\sigma}\!\!c_{i\sigma}^\dagger c_{j\sigma}
+\!\!\sum_i U_i\hat{n}_{i\uparrow} \hat{n}_{i\downarrow}\!
+\!\!\sum_{i,\sigma} \epsilon_i\hat{n}_{i\sigma}\!+\!\hat{W}(\tau),
\label{Hamil}
\end{align}
where $\langle ij\rangle$ denotes nearest neighbor sites, $\sigma=\uparrow,\downarrow$ and 
$\hat{n}_{i\sigma}=c_{i\sigma}^\dagger c_{i\sigma}$ is the local density operator. We take $t=1$ as energy unit.
The subscript $i$ in the onsite energy $\epsilon_i$ and repulsion term $U_i$ allows
for possible inhomogeneities. The external potential in time-dependent calculations is
$\hat{W}(\tau)=\sum_{i\sigma}w_i(\tau)\hat{n}_{i\sigma}$, with $\tau$ being the time variable.
%

{\it Dynamical Mean Field Theory (DMFT).-}
 In this study we neglect magnetic phases and allow only for paramagnetic solutions ($n_\uparrow=n_\downarrow$), to 
focus on pure Mott physics. DMFT maps a Hubbard model on a simple 
cubic lattice onto a local problem representing one of the lattice sites (site 0)
surrounded by a bath which describes the rest of the lattice \cite{revdmft}. 
In practice, one introduces auxiliary degrees of freedom to recover a Hamiltonian description
 of the local problem by identifying the site 0 with the impurity site of an Anderson impurity model
(AIM): 
\begin{equation}
\label{aim}
\!{\cal{ H}}_{AIM}  = \sum_{l,\sigma}
\left[ \epsilon_l a^{\dag}_{l\sigma} \, a_{l\sigma} + V_l\,
(a^{\dag}_{l\sigma} c_{0\sigma} + \mbox{h.c.}) \right]+{\cal{H}}_{imp},
\end{equation}
where $ {\cal{H}}_{imp} = Un_{0\uparrow} n_{0\downarrow} - \mu n_0$, 
 $\mu$ is  the chemical potential,
and the parameters $V_l,\epsilon_l$ are determined self-consistently.
 Self-consistency with the original lattice is obtained by requiring the impurity single-particle Green function 
(in Matsubara space \cite{definitions}) $G(i\omega_n)$ to be identical to the local
lattice Green function with identical self-energy $\Sigma(i\omega_n)$, i.e.
\begin{equation}
 G(i \omega_n) = \int d\epsilon  D(\epsilon)[i\omega_n+\mu-\epsilon-\Sigma(i\omega_n)]^{-1}
\end{equation}
where $\Sigma(i\omega_n)$ is obtained via the local Dyson equation and
$D(\epsilon)$ is the non-interacting lattice density of states.

We solved the self-consistent AIM using the exact-diagonalization (Lanczos) algorithm \cite{krauth},
which truncates the number of auxiliary degrees of freedom to a finite, small number $N_s$.
The results shown $(N_s=8)$ are converged against $N_s$. 
Once at self-consistency, the density $n=\sum_\sigma \langle n_{0\sigma} \rangle$,
 the average double occupancy $d= \langle n_{0\uparrow}n_{0\downarrow} \rangle$
 (and thus the potential energy per lattice site $V=Ud$) are evaluated as averages
 on the impurity site ($0$) of the AIM. The total energy per site is given by $E_{DMFT}=K_{DMFT}+V$, where $K_{DMFT}$ is given by \cite{definitions}
\begin{equation}
 K_{DMFT}=\frac{2}{\beta}\sum_n e^{i\omega_n 0^+}\!\!\int d\epsilon \frac{\epsilon D(\epsilon)\ }{i\omega_n+\mu-\epsilon-\Sigma(i\omega_n)} 
\end{equation}
\begin{figure}[b]
\includegraphics[width=75mm,angle=-0]{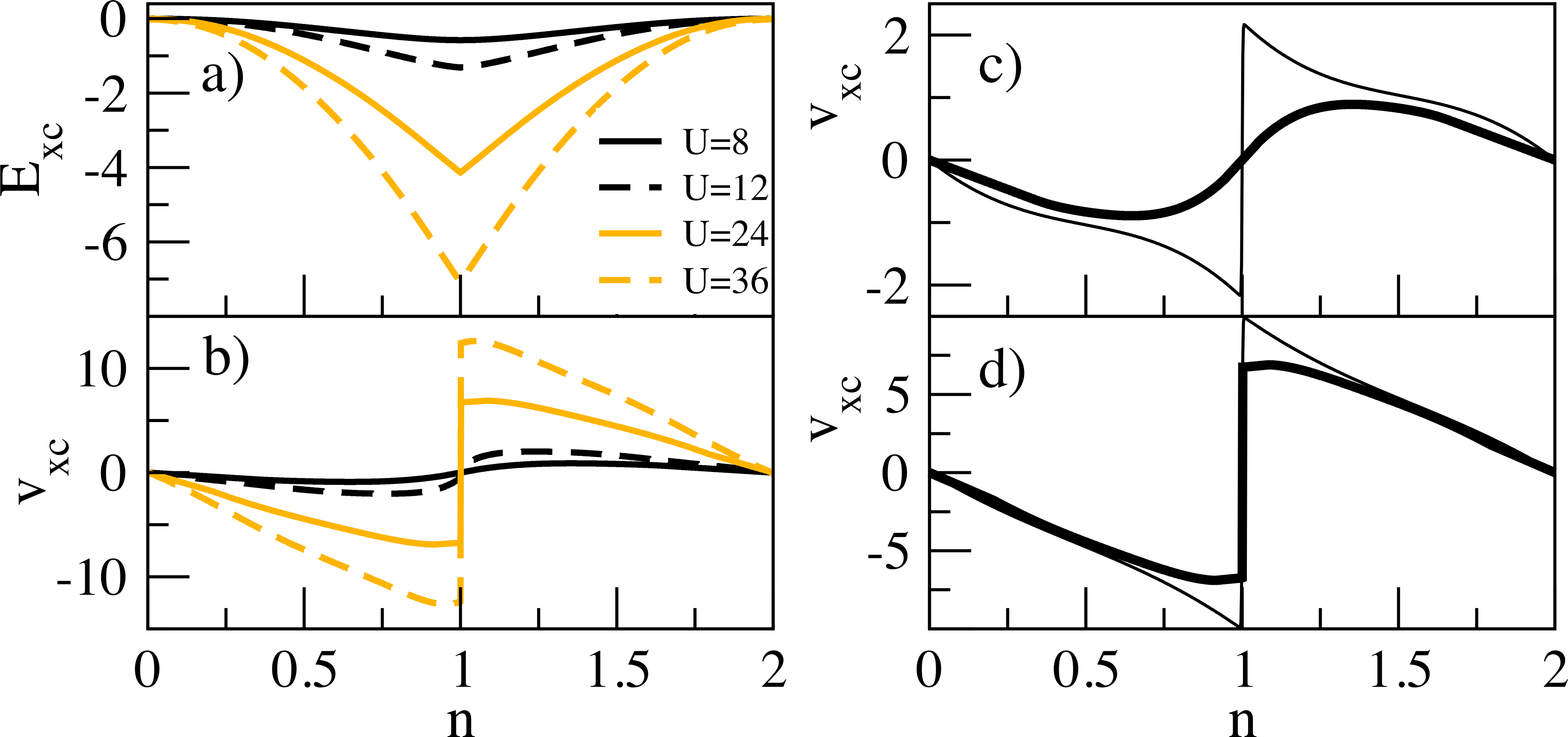}
\caption{(Color online) a-b): Exchange-correlation energies $E_{xc}$ and potentials $v_{xc}$
for the homogeneous 3D Hubbard model, for several values of the interaction $U$.
c): DMFT (thick solid curve) vs 1D BALDA results (thin solid curve) for $v_{xc}$ when $U=8$. d): 
same as c), but $U=24$. }
\label{Fig1} 
\end{figure}
%
%
%
{\it DFT for the 3D Hubbard model.-}  In a spin-independent DFT for the Hubbard model \cite{GunSchon},
the total energy is:
\begin{align}
E_v [n] \equiv T_0[n] + E_H[n] + E_{xc}[n]+\sum_i v_{ext}(i) n_i ,
\label{theory::Exc}
\end{align}
where $v_{ext}$ denotes the static external field (in the notation of Eq. (\ref{Hamil}),
$v_{ext}(i)\equiv \epsilon_i$). In Eq. (\ref{theory::Exc})
$n_i=\sum_\sigma n_{i\sigma}$,  while $T_0[n]$ and $E_H = \frac{1}{4} \sum_i U_i n_i^2$ are, respectively, 
the kinetic energy of the non-interacting system  and the Hartree energy.

We use an LDA for $E_{xc}$ and $v_{xc}$: $v_{xc}(i) = v_{xc} (n_i)$,
where $E_{xc}$ is obtained from the
homogeneous 3D Hubbard model, our reference system. We employed DMFT to obtain
$E_{xc} = E_{DMFT} - T_0 - E_H$, with $E_{DMFT}$ being the ground state energy of the reference system. 
The DMFT impurity solver introduced some noise in the numerical solution, especially 
at low densities ($n <  0.2$) or close to half-filling ($n \lesssim 1$). 
Thus we first smoothed the data, and then performed a polynomial fitting for $0.2 \le n \le 1$.
For $n <0.2$,  instead of DMFT, we used an analytic, asymptotically 
exact, form for the ground state energy of the 3D Hubbard model \cite{Giuliani}.
Including the sub-leading term at low $n$, one gets 
$E_{xc} = \left ( 8 \pi a_s^U - U \right ) n^2 / 4 + \lambda n^{7/3}$, where 
$a_s^U = \frac{1}{8 \pi} \frac{1}{U^{-1}+\gamma}$ 
is the scattering length for the model,
$\gamma = 0.1263t^{-1}$ for a simple cubic lattice, and $ \lambda$ is a fitting parameter.
The piecewise analytical expression for $E_{xc}$ was then
differentiated to obtain $v_*(n)$, the exchange-correlation potential for $0\leq n<1$. Due
to electron-hole symmetry, in the entire density range $[0,2]$, we have $v_{xc}^{DMFT}=\theta(1-n)v_*(n)-\theta (n-1)v_*(2-n)$,
where $\theta$ is the step function. 

Results for $E_{xc}$ and $v_{xc}$ from DMFT are in Figs. \ref{Fig1}a-b
for several $U$ values. On increasing $U$, a change of curvature occurs in $E_{xc}$ for
$n \approx 1$ (Fig. \ref{Fig1}a), and a cusp develops above a critical
value $U_c^{Mott} \approx 14$. This induces a discontinuity in $v_{xc}$ (Fig. \ref{Fig1}b), 
a manifestation of the Mott-Hubbard metal-insulator transition in a DFT description.
Such behavior is quite different from that  of $v_{xc}$ in the 1D Hubbard model, where
the discontinuity occurs for any $U>0$  \cite{Capelle03}.
In Figs. \ref{Fig1}c-d
we present results for $v_{xc}^{DMFT}$ (present work) and $v_{xc}^{BALDA}$ \cite{Capelle03} for $U=8$ and 24.
For $U=8$, only $v^{BALDA}_{xc}$ is discontinuous. 
The DMFT and BALDA exchange-correlation potentials are rather different from each other in the entire density
range, with $|v_{xc}^{BALDA}| > |v_{xc}^{DMFT}|$, reflecting the difference between 1D and 3D reference systems. 
These features are generic for any $U$ value.
In particular, for $U>U_{c}^{Mott}$, the discontinuity in $v_{xc}^{BALDA}$ is considerably 
larger than in $v_{xc}^{DMFT}$, as seen in Fig. \ref{Fig1}d. 

To ease the numerics, we slightly smoothed near $n=1$
the $v_{xc}$:s for $U>U_c^{Mott}$. 
The $v^{DMFT}_{xc}$ thus obtained was used in our initial, ground state DFT-LDA calculations,
via the Kohn-Sham (KS) equations 
\begin{equation}
( \hat{T} + \hat{v}_{KS} ) \varphi _\kappa = \varepsilon_\kappa\varphi _\kappa \ \ ,                                 
\end{equation}
 where $\hat{T}= -t\sum_{\langle ij\rangle,\sigma} c_{i\sigma}^\dagger c_{j\sigma}$ and $\varphi_\kappa$
is the $\kappa$-th single particle KS orbital, with $n_i = \sum^{occ} _\kappa |\varphi _\kappa (i)|^2$.
The effective potential $v_{KS} (i) = v_H(i) + v_{xc} (i) + v_{ext}(i)$, with  $v_H(i)=\frac{1}{2} U_i n_i$
the Hartree potential, and $v_{xc}(i) = v^{DMFT}_{xc}(n_i)$.

%
%
{\it TDDFT for the 3D Hubbard model. -} 
To perform TDDFT real-time dynamics of the Hubbard model \cite{Verdozzi08},
one propagates in time the KS orbitals $\varphi _\kappa(\tau)$ via the
time-dependent Kohn-Sham equations: 
\begin{equation}
( \hat{T} + \hat{v}_{KS}(\tau) ) \varphi _\kappa(\tau) = i  \partial_\tau \varphi _\kappa(\tau)\ ,
\end{equation}
to get the density $ n_i(\tau) = \sum^{occ} _{\kappa} |\varphi _\kappa (i,\tau)|^2$.
In general, $ v_{KS} (i,\tau) = v_H(i,\tau) + v_{xc} (i,\tau) + v_{ext}(i,\tau)$ depends non-locally 
on the density via $v_{xc}$. In the adiabatic LDA considered below,
a local dependence in space and time is assumed: $ v_{xc}( i, \tau) \rightarrow v^{DMFT}_{xc} (n_i(\tau))$.\newline
\begin{figure}[b]
\includegraphics[width=87mm,angle=-0]{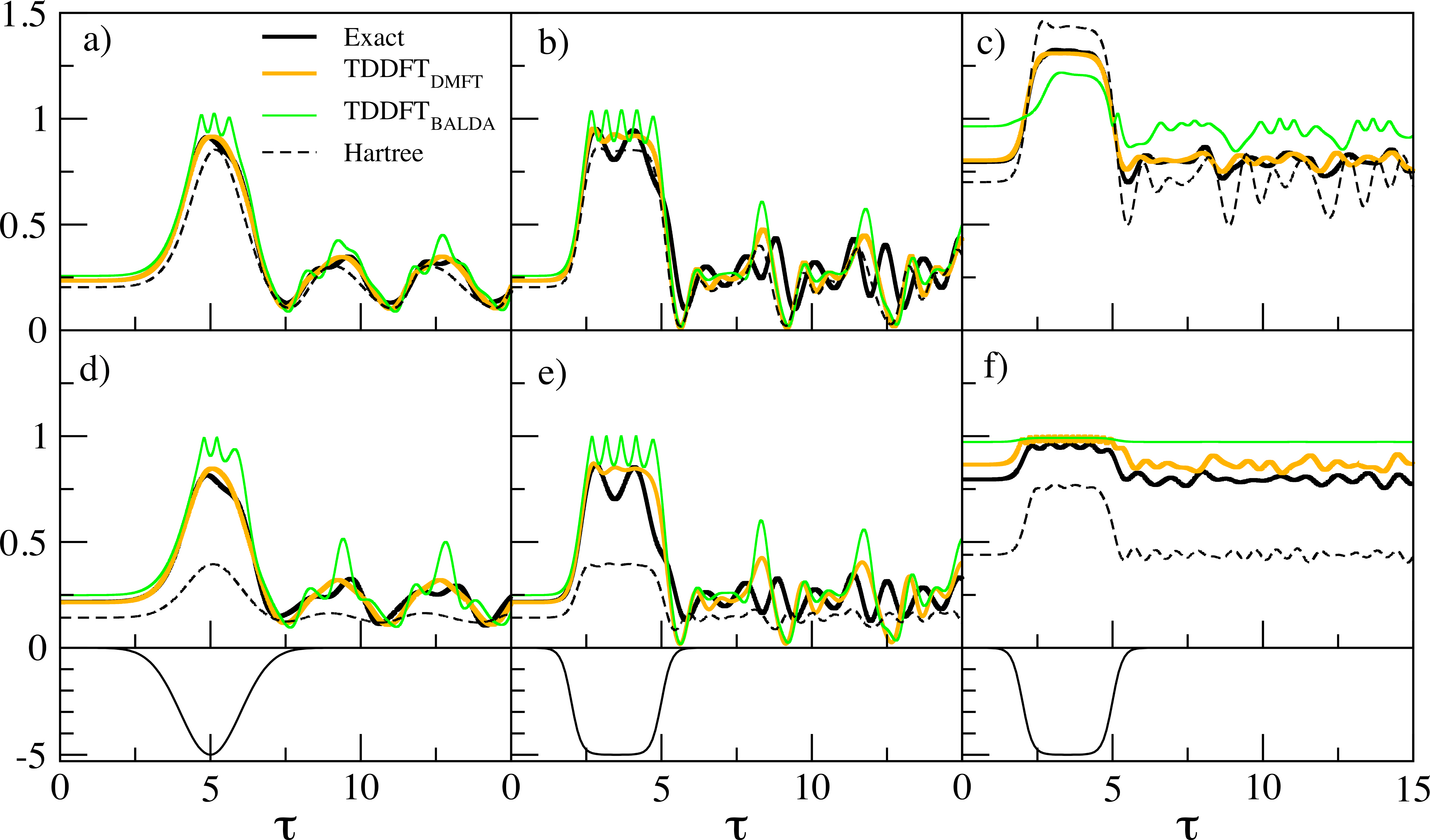}
\caption{(Color online) Exact, TDDFT and Hartree density  for $U=8$ (a-c) and $U=24$ (d-f)
at the central site of a $5\times5\times5$ Anderson impurity cluster. 
In (c) and (f) the cluster has 
$N_e=70$ electrons, and  $\epsilon_0=-2.66$ and $\epsilon_0=-4$, respectively, to attain an initial (and the same)
density close to half-filling. Otherwise $N_e=40$ \cite{special} and  $\epsilon_0=0$.
The time-dependent perturbation $w_0(\tau)$ acts always only at the impurity site, and its shape is shown below (d-f) (see main text for details).}
\label{Fig2} 
\end{figure}
%
%
{\it Exact vs adiabatic-LDA dynamics. -} 
In Fig. \ref{Fig2} we compare the $A^{LDA}_{DMFT}$ densities  to the exact ones,
for a simple cubic cluster with $5^3$ sites and open boundary conditions. We consider
a highly inhomogeneous case, a single interacting impurity in the center $i=0$:
$U_i=U\delta_{i0}$ in Eq.(\ref{Hamil}), which
should be a rather severe test for an adiabatic LDA based on a DMFT of the homogeneous 3D Hubbard model. 
We also set $\epsilon_i=\epsilon_0 \delta_{i0}$ and $w_i(\tau)=w_0(\tau)\delta_{i0}$.
Due to symmetry, only 10 ($N_{sy}$) out of the total (per spin)  $125$ one-particle eigenstates, those with 
nonzero  amplitude at $i=0$ (active states), determine 
the static and time-dependent density at $i=0$, making the size of the exact configuration space
manageable \cite{special}. 

We considered two time profiles for $w_0(\tau)$: gaussian
and rectangle-shaped. In the latter, the ramping up/down of the pulse is faster. 
For $U=8$, Fig. \ref {Fig2}a-c, there is a very good agreement between
$A^{LDA}_{DMFT}$ and the exact results when the perturbation is actually present.
Afterwards, when $w_0(\tau)$ has returned to zero,
the agreement somewhat deteriorates for the faster perturbation, indicating 
the presence of non-adiabatic, non-local effects in the exact dynamics. 
These however are not described by $A^{LDA}_{DMFT}$.  
Similar considerations apply for $U=24$, Fig.\ref {Fig2}d-f. Here,
$v_{xc}^{DMFT}$ exhibits a gap, and yet the $A^{LDA}_{DMFT}$ performs rather well at low filling (panels d-e).
The performance worsens considerably closer to half-filling, panel f),  for two
reasons. 

The first concerns the ground state densities in Fig. \ref{Fig2} (i.e. at $\tau=0$).
When obtained with $v_{xc}^{DMFT}$, these agree very well with the exact ones, save for panel f).
To understand why, we determined via reverse-engineering the exact ground state 
Kohn-Sham potential $v^{Ex.}_{KS}$ for all cases a-f).
For panels a-e), $v^{Ex.}_{KS}\approx 0$ for $i\neq 0$,
i.e. $v_{xc}$ is essentially local. For case f), we found large nonzero values of $v^{Ex.}_{KS}$
at $i\neq 0$. Such nonlocality is missing in our $A^{LDA}_{DMFT}$. The second reason is
that in Fig. \ref{Fig2}f the $A^{LDA}_{DMFT}$ density crosses the discontinuity in the $v_{xc}$ 
at half-filling. Such discontinuity was determined from the infinite, homogeneous 3D Hubbard model,  but is
expected to be significantly modified in an inhomogeneous small cluster: such change is missed 
by our $v_{xc}^{DMFT}$. This causes the disagreement in Fig. \ref{Fig2}f (e.g.,
oscillations occur around $n_0=1$ in the $A^{LDA}_{DMFT}$ densities, which are different from those
in the exact curve). 

In Fig. \ref{Fig2} we also display results obtained with $v_{xc}^{BALDA}$ and $v_{xc}=0$ (Hartree dynamics).
While conceptually unjustified in 3D, the adiabatic BALDA is an helpful tool to asses the validity
of $A^{LDA}_{DMFT}$. Thereby we see that for $U=8$ the adiabatic BALDA induces density oscillations due to 
the discontinuity at half-filling which, in $v_{xc}^{BALDA}$ (but not in $v_{xc}^{DMFT}$) exists for any $U>0$.
Also, $v_{xc}^{BALDA}$
is considerably stronger than the DMFT counterpart, see e.g. Fig. \ref{Fig2}f: The adiabatic BALDA 
density does not cross $n_0=1$, due to the too large discontinuity of  $v_{xc}^{BALDA}$.
This shows that a $v_{xc}$ which 
properly includes correlations in 3D is needed.
Finally, the Hartree dynamics is
much worse than the $A^{LDA}_{DMFT}$ one. Other densities and perturbations
gave consistent results \cite{MasterDaniel}.\newline
%
%
%
\indent {\it Bloch Oscillations. -}
As an application of $A^{LDA}_{DMFT}$, we
discuss briefly  the Bloch oscillations in the 3D Hubbard model. 
The Bloch oscillations are the response of particles in a lattice
to an electric field $F$ \cite{Hartmann_NJP_6_2}. 
With no interactions, in a lattice a uniform $F$ linearly increases the particle momentum until Bragg reflections occur and
an oscillatory current sets in with frequency $\omega = F$, i.e. the Bloch oscillations. These 
have been observed in superlattices and, quite recently, in ultracold-atom systems \cite{blochexp}. 
The effect of strong interactions on the Bloch oscillations has been theoretically investigated
mostly for 1D and 2D bosons, and for 1D and infinite-dimensional fermions \cite{prl1,Freericks}: 
Depending on the Hamiltonian parameters ($U/t$  and $F/t$ in Hubbard models) damped behavior, beatings, 
or highly irregular patterns can take place. Here, in analogy with cold-atom experimental setups, we consider 
a 3D Hubbard model with an asymmetric parabolic confinement. 
We use a 3D cluster of $33\times 5 \times 5$ sites (along $x,y,z$) with open boundary conditions.  
To have a ground state atomic "cloud" elongated in the $x$ direction,
we take a potential $ \epsilon_i = \frac{1}{2} \left [ k_x x_i^2 + k_{yz} (y_i^2 + z_i^2) \right ]$ with $k_x < k_{yz}$.
\newline
\begin{figure}[t]
 \includegraphics[width=0.45\textwidth,angle=-0]{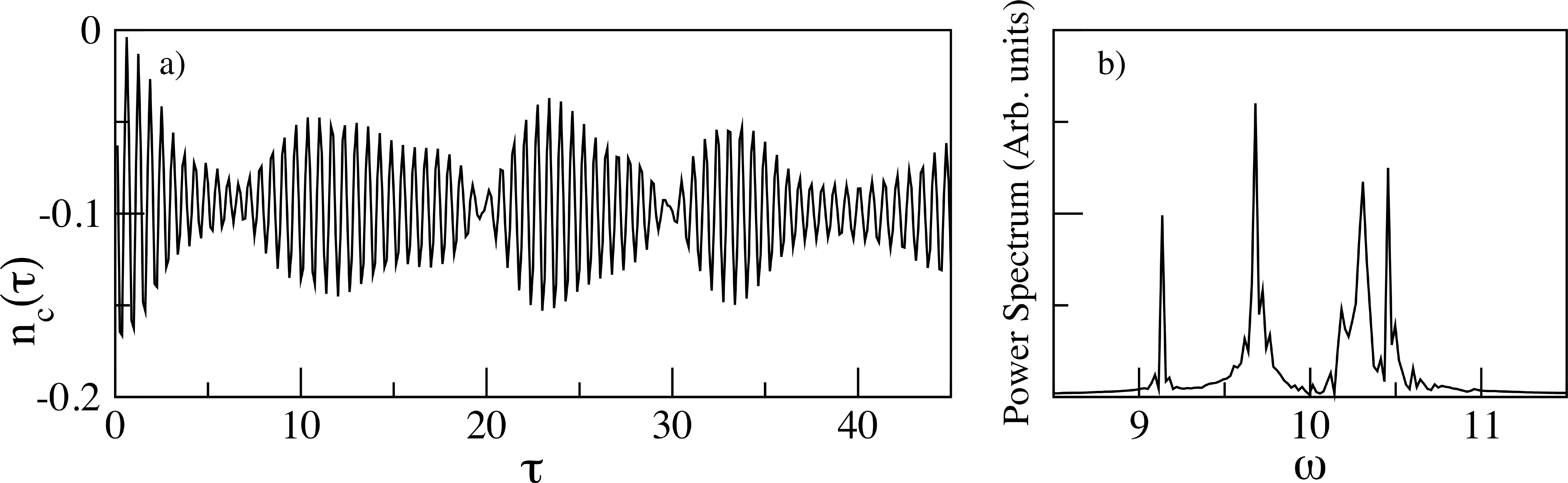}
 \vspace{-2mm}
\caption{\label{damping} Beats regime, with $U=2, N_\uparrow=N_\downarrow=8, F=10, k_x=1,k_{yz}=10$. 
a) Bloch oscillations of the density 
centroid $n_c(\tau)$ b) Power spectrum of $n_c$. The temporal patterns persist for the rest
of the (larger than in figure) simulation interval.
}
\vspace{-2mm}
\label{Fig3}    
\end{figure}
At $\tau=0^+$, we start the time evolution with a perturbation $w_i(\tau ) =\theta(\tau)\left[- \frac{1}{2} k_x x_i^2  + F x_i \right]$.
Our time-dependent indicator for the Bloch oscillations is the one-particle centroid
$ n_c(\tau) = N^{-1}_p \sum_i (x_i-x_0) n_i(\tau)$,
with $n_i$ the particle density at site $i$, $N_p =\sum_i n_i$,
and $x_0$ the center of the cluster in the $x$-direction.
We examined several regimes where beats and damping of the oscillations 
are expected (the inherent parameter values were chosen by adapting to 3D the corresponding ones for
such regimes in 1D). For large driving fields, we found beats in the $A^{LDA}_{DMFT}$
results (Fig. \ref{Fig3}a), with a frequency $\omega \approx F$ and extra peaks with a 
splitting $\Delta\omega \approx U$
(Fig. \ref{Fig3}b). No unequivocal signatures of damped oscillations were observed
for several setups in the expected parameter regime:
It is highly plausible that this is due to the lack of memory in the $A^{LDA}_{DMFT}$,
suggesting the need for improved, non-adiabatic exchange-correlations potentials. \newline
%
%
%
\indent {\it Conclusions. -} We introduced a method to determine new exchange-correlation
potentials for inhomogeneous strongly correlated systems in 3D. 
We used DMFT as "engine" for  the many-body calculations, but one could
equally well have considered other approaches, such as, e.g., the Gutzwiller approximation or the Quantum Monte Carlo method.
For 3D clusters with Hubbard-type interactions, {$A^{LDA}_{DMFT}$ and exact results
are in good agreement. While, in general, both nonlocal and non-adiabatic effects are needed in $v_{xc}$,
we have here shown that an $A^{LDA}_{DMFT}$, i.e. a $v_{xc}$ which correctly incorporates 
correlations in 3D,  is an effective starting point for a TDDFT description of strongly
correlated systems in 3D and out of equilibrium. Moreover, by involving only one time variable,
our approach offers a clear computational advantage in describing the time evolution
of large inhomogeneous 3D systems. Hopefully, these attractive features will stimulate further studies and pave the way to efficient and accurate [TD]DFT  
treatments of the equilibrium and nonequilibrium behavior of 3D strongly correlated materials.\newline
We  thank C.-O. Almbladh and U. von Barth for discussions. C.V. is supported by ETSF(INFRA-2007-211956).
\vspace{-2mm}
%
%


\begin{thebibliography}{9}
\bibitem{definitions} $G(i\omega_n)$ is the Matsubara transform of  $G(\tau)= - \langle \mathcal{T} \{c_{0 \sigma} (\tau) c^\dagger_{0 \sigma}\}\rangle_{AIM}$,
 where $\mathcal{T}$ is the time ordering operator, $\omega_n=(2n+1)\pi/\beta$, and $\beta^{-1}=T$ ($k_B=1$).
 To obtain the ground state energy, we let $\beta \to \infty$.
\end{thebibliography}
\end{document}